\documentclass{article}
\usepackage[T1]{fontenc}
\usepackage{wrapfig}
\usepackage[portuges,english]{babel}
\usepackage{amssymb,latexsym,amsmath,color,mathrsfs,ifsym,graphics,stmaryrd} %stmaryrd for \llbracket and \rrbracket
\usepackage[colorlinks,linkcolor=blue,urlcolor=blue,citecolor=black,
plainpages=false,pdfpagelabels,breaklinks]{hyperref}

\title{Immanent Powers versus Causal Powers\\ (Propensities, Latencies and Dispositions)\\ in Quantum Mechanics}
\author{{\sc Christian de Ronde}\thanks{Fellow Researcher of the Consejo
Nacional de Investigaciones Cient\'{\i}ficas y T\'ecnicas and Adjoint Professor of the National University Arturo Jaurteche.}}
\date{\begin{center}
\begin{small}
Philosophy Institute Dr. A. Korn \\ 
Buenos Aires University, CONICET - Argentina \\
Center Leo Apostel and Foundations of  the Exact Sciences\\
Brussels Free University - Belgium \\
\end{small}
\end{center}}

\usepackage[margin=2.5cm]{geometry}

\begin{document}

\maketitle

\bigskip

\bigskip

\begin{abstract}
\noindent In this paper we compare two different notions of `power', both of which attempt to provide a realist understanding of quantum mechanics grounded on the potential mode of existence. For this propose we will begin by introducing two different notions of potentiality present already within Aristotelian metaphysics, namely, {\it irrational potentiality} and {\it rational potentiality}. After discussing the role played by potentiality within classical and quantum mechanics, we will address the notion of {\it causal power} which is directly related to irrational potentiality and has been adopted by many interpretations of QM. We will then present the notion of {\it immanent power} which relates to rational potentiality and argue that this new concept presents important advantages regarding the possibilities it provides for understanding in a novel manner the theory of quanta. We end our paper with a comparison between both notions of `power', stressing some radical differences between them. 
\end{abstract}

\bigskip


%--------------------------------------------------------------
\renewenvironment{enumerate}{\begin{list}{}{\rm \labelwidth 0mm
\leftmargin 0mm}} {\end{list}}

\newcommand{\ita}{\textit}
\newcommand{\mcal}{\mathcal}
\newcommand{\mfrak}{\mathfrak}
\newcommand{\mbb}{\mathbb}
\newcommand{\mrm}{\mathrm}
\newcommand{\msf}{\mathsf}
\newcommand{\mscr}{\mathscr}
\newcommand{\lra}{\leftrightarrow}
\renewenvironment{enumerate}{\begin{list}{}{\rm \labelwidth 0mm
\leftmargin 5mm}} {\end{list}}

\newtheorem{dfn}{\sc{Definition}}[section]
\newtheorem{thm}{\sc{Theorem}}[section]
\newtheorem{lem}{\sc{Lemma}}[section]
\newtheorem{cor}[thm]{\sc{Corollary}}
\newcommand{\Proof}{\textit{Proof:} \,}
\newcommand{\cqd}{{\rule{.70ex}{2ex}} \medskip}

\bigskip

\bigskip

\bigskip

\section*{Introduction}

The notion of potentiality has played a major role within the history of quantum physics. Its explicit introduction within the theory goes back to the late 1950' when several authors discussed ---independently--- the possibility to understand Quantum Mechanics (QM), in close analogy to the Aristotelian hylomorphic metaphysical scheme, through the consideration of a potential realm ---different to that of actuality. Werner Heisenberg \cite{Heis58}, Gilbert Simondon \cite{Simondon58}, Henry Margenau \cite{Margenau54}, and Karl Popper \cite{Popper59}, presented different interpretations of QM in terms of potentialities, propensities and latencies. Since then, these interpretations and ideas have been developed in different directions enriching the debate about the meaning and possibilities of the theory \cite{Aerts81, Cartwright89, Cartwright07, Dorato06, DoratoEsfeld, Esfeld11, Piron76, Piron81, Suarez07}. However, as we shall argue, most of these attempts ground themselves on a one sided view of potentiality, one which understands potentiality as defined exclusively in terms of actual effectuations ---i.e., in terms of what Aristotle called {\it irrational potentiality}. In particular, dispositional and propensity type interpretations of QM have been linked through this understanding of potentiality to the concept of {\it causal power}. In this paper ---continuing with ongoing work \cite{deRonde13, deRonde16a, deRonde17}--- we investigate a different standpoint, one which attempts to develop potentiality on the lines of {\it rational potentiality}.\footnote{We might also consider some interpretations of QM (e.g., \cite{Aerts10, Kastner12, Karakostas07}) as closer to the development of rational potentiality presented here.} In turn, this new potential mode of existence can be related to the physical concept of {\it immanent power}. We believe that this new understanding of the meaning of both the `potential realm' and the notion of `power' can provide us with new insights that might help us to understand what QM is really talking about. 

The paper is organized as follows. In the first section we outline the main aspects of the orthodox empiricist project within philosophy of QM. In section 2 we present the representational realist project which attempts to produce a different account of the problem of interpreting quantum theory. Section 3 recalls the metaphysical hylomorphic Aristotelian scheme and the distinction between, on the one hand, actuality and potential modes of existence, and on the other, irrational and rational potentiality. In section 4, we analyze the Newtonian atomistic representation which eliminated the potential realm from the description of physical reality. In section 5, we discuss Heisenberg's return to hylomorphic metaphysics ---through the reintroduction of the potential realm--- in order to interpret QM. Section 6 presents the continuation of Heisenberg's proposal following the teleological scheme of causal powers, dispositions, propensities, etc. ---all of which are based on the notion of irrational potentiality. In section 7 we present a different non-reductionistic scheme based on the notion of rational potentiality which attempts to discuss the conceptual representation of the potential realm in terms of the notion of immanent power. Section 8 analyses and discusses the pros and cons of the notions of causal power and immanent power in order to provide an answer to the question: what is QM really talking about?

\section{The Orthodox (Empiricist) Project in Philosophy of QM}

Philosophy of science in general, and philosophy of physics in particular, were developed after the second world war following the logical positivist project proposed by the so called ``Vienna Circle''. Following the physicist and philosopher Ernst Mach, logical positivists fought strongly against dogmatic metaphysical thought and {\it a priori} concepts. As they argued in their famous {\it Manifesto} \cite{VC}: ``Everything is accessible to man; and man is the measure of all  things. Here is an affinity with the Sophists, not with the Platonists; with the Epicureans, not with the Pythagoreans; with all those who stand for earthly being and the here and now.'' Their main attack against metaphysics was developed taking an empiricist based standpoint, the idea that one should focus in ``statements as they are made by empirical science; their meaning can be determined by logical analysis or, more precisely, through reduction to the simplest statements about the empirically given.'' The positivist architectonic stood on the distinction between {\it empirical terms}, the empirically ``given'' through observation,\footnote{Later on considered as {\it observational terms}.} and {\it theoretical terms}, their translation into simple statements. This separation and correspondence between theoretical statements and empirical observation would have deep consequences, not only regarding the problems addressed within the new born ``philosophy of science'' but also with respect to the limits in the development of many different lines of research within the theory of quanta itself.

The main enemy of empiricism has been, since its origin, metaphysical thought ---understood mainly as a discourse about non-observable entities. Trying to avoid any metaphysical reference beyond observational phenomena, the empiricist perspective attempted to produce a direct link between observation on the one hand, and linguistic statements on the other. But how to do this without entering the field of metaphysical speculation, specially when language and metaphysics are intrinsically related? Indeed, (metaphysical) concepts, as a necessary prerequisite to account for experience, are defined ---since Plato and Aristotle--- in a systematic and categorical manner; i.e. in a metaphysical fashion. In this respect, we remark that metaphysics has nothing to do with the distinction between observable and non-observable; it deals instead with the very possibility of defining concepts systematically through general principles.  

The main problem of empiricism has been clearly exposed by Jorge Luis Borges in a beautiful short story called {\it Funes the Memorious} \cite{Borges}. Borges recalls his encounter with Ireneo Funes, a young man from Fray Bentos who after having an accident become paralyzed. Since then Funes' perception and memory became infallible. According to Borges, the least important of his recollections was more minutely precise and more lively than our perception of a physical pleasure or a physical torment. However, as Borges also remarked: ``He was, let us not forget, almost incapable of general, platonic ideas. It was not only difficult for him to understand that the generic term dog embraced so many unlike specimens of differing sizes and different forms; he was disturbed by the fact that a dog at three-fourteen (seen in profile) should have the same name as the dog at three fifteen (seen from the front). [...] Without effort, he had learned English, French, Portuguese, Latin. I suspect, however, that he was not very capable of thought. To think is to forget differences, generalize, make abstractions. In the teeming world of Funes there were only details, almost immediate in their presence.'' The problem exposed by Borges is in fact, the same problem which Carnap \cite{Carnap28}, Nagel \cite{Nagel61}, Popper \cite{Popper35} and many others ---following the positivist agenda--- tried to resolve: the difficult relation between, on the one hand, phenomenological experience or observations, and on the other, language and concepts.  

The failure of the orthodox empiricist project to define {\it empirical terms} independently of theoretical and metaphysical considerations was soon acknowledged within philosophy of science itself. In the sixties and seventies important authors within the field ---such as Hanson, Kuhn, Lakatos and Feyerabend, between many others--- addressed the problem of understanding observation in ``naive'' terms. This debate, which unmasked the limits of the empiricist project, was known to the community by the name of: ``the theory ladenness of physical observation''. But regardless of the impossibility to consider observation as a ``self evident given''  ---as the young born positivist epistemology had attempted to do---, philosophy of science in general, and philosophy of physics in particular, have continued anyhow to ground their analysis in ``common sense'' observability. Indeed, as remarked by Curd and Cover \cite[p. 1228]{PS}: ``Logical positivism is dead and logical empiricism is no longer an avowed school of philosophical thought. But despite our historical and philosophical distance from logical positivism and empiricism, their influence can be felt. An important part of their legacy is observational-theoretical distinction itself, which continues to play a central role in debates about scientific realism.'' It is important to notice that the orthodox problems of QM have been also constrained by the empiricist viewpoint which understands that observation in the lab, even in the case of considering quantum phenomena, remains completely unproblematic. 

Within the huge literature regarding the meaning of QM, the empiricist standpoint has constrained philosophical analysis within the walls of a conservative project which attempts to ``bridge the gap'' between the ``weird''  mathematical formalism of QM and our ``manifest image of the world'' \cite{Dorato15}. Indeed, going back to Bohr's reductionistic desiderata,\footnote{We refer to the Bhorian presuppositions according to which, first, QM must be related through a limit to classical physics, and second, experience and phenomena will be represented always in terms of classical physics. See for a detailed discussion and analysis \cite{deRonde16b}.} the main goal of the orthodox project in philosophy of QM is to understand the theory in strict relation to our classical ``common sense'' representation of the world. This reductionistic account of QM is the reason why the quantum to classical limit has been considered within the literature as one of the most important problems. The failure to solve this problem through ``the new orthodoxy of decoherence'' \cite[p. 212]{Bub97} remains still today partly unnoticed by the community and even camouflaged through a ``FAPP (For All Practical Purposes) justification'' which confuses the epistemological and ontological levels of analysis \cite{deRonde16c}. In this same context, the orthodox perspective has created a set of ``no-problems'' (non-separability, non-individuality, non-locality, non-distributivity, non-identity, etc.) which discuss the quantum formalism presupposing ``right from the start'' the classical notions with which we have been able to build our classical (metaphysical) representation of the world (see for a detailed discussion \cite{deRonde16b}). This ``conservative project'' within philosophy of QM has silenced a radically different line of research which would investigate the possibilities of a truly non-classical metaphysical representation of QM.

\section{The Representational Realist Project} 

Representational realism understands physics as a discipline which attempts to represent {\it physis} (reality or nature) in theoretical ---both formal and conceptual--- terms  \cite{Cordero}. Physical theories provide, through the tight inter-relation of mathematical formalisms and networks of physical concepts, the possibility of representing experience and reality. In this respect, we must stress that the possibility to imagine and picture reality beyond observation is only provided through physical concepts, not through mathematical formalisms. Mathematics is an abstract discipline which contains no physical concept whatsoever. And this is the reason why mathematicians can work perfectly well without learning about physical theories or about the relation between certain mathematical formalisms and physical reality. Physical concepts cannot be ``found'' or ``discovered''  within any mathematical formalism. The theory of calculus does not contain the Newtonian notions of absolute `space' and `time', it does not talk about `force', `particles', `mass' or `gravity'. The physical notion of field cannot be derived through a theorem from Maxwell's equations. Physical concepts are not mathematical entities, they are metaphysical elements created in systematic categorical terms through general principles ---such as, for example, those proposed by Aristotle in his metaphysics and logic. 

According to the representational realist, reality is not something ``self-evidently'' exposed through observations ---as logico-positivists, empricists and even Bohr might have claimed---, but on the contrary, its representation and understanding is only specified through the metaphysical conceptual scheme provided by physical theories themselves. To avoid any misunderstanding, let us stress that our realist viewpoint is not consistent with scientific realism, phenomenological realism or realism about observables ---which we consider to be in fact variants of empiricism grounded on ``common sense'' observability. As Musgrave \cite[p. 1221]{PS} makes the point: ``In traditional discussions of scientific realism, common sense realism regarding tables and chairs (or the moon) is accepted as unproblematic by both sides. Attention is focused on the difficulties of scientific realism regarding `unobservables' like electrons.'' Contrary to the positivist viewpoint which assumes as a standpoint the ``common sense'' observability of tables and chairs, the goal of physics has been always ---since its origin in Greece around the VII century B.C.--- the theoretical representation of {\it physis} \cite{Cordero}. This scientific project has nothing to do with ``common sense''. As Heraclitus himself remarked many centuries ago: {\it physis} loves to hide. 

According to our viewpoint, there is no physical observation without the aid of a network of adequate concepts. It is important to stress at this point that we use the term `metaphysics' to refer to the systematic definition of conceptual schemes. A metaphysical scheme is nothing more, and nothing else, than a conceptual net of interrelated concepts. It provides the very preconditions of observability itself. As remarked by Borges, without concepts, generalizations which escape differences, particular experiences, it is not possible to provide meaning to experience, it is not possible to think. From our perspective metaphysics has nothing to do with the the empiricist based distinction between observable entities (e.g., tables and chairs) and un-observable entities (e.g., atoms). Following Tian Yu Cao, we might regard this distinction as ineffective to account for the present challenges within philosophy of QM: 

\begin{quotation}
\noindent {\small``The old-fashioned (positivist or constructive empiricist) tradition to the distinction between observable and unobservable entities is obsolete. In the context of modern physics, the distinction that really matters is whether or not an entity is cognitively accessible by means of experimental equipment as well as conceptual, theoretical and mathematical apparatus. If a microscopic entity, such as a W-boson, is cognitively accessible, then it is not that different from a table or a chair. It is clear that the old constructive empiricist distinction between observables and nonobservables is simply impotent in addressing contemporary scientific endeavor, and thus carries no weight at all. If, however, some metaphysical category of microscopic entities is cognitively inaccessible in modern physics, then, no matter how basic it was in traditional metaphysics, it is irrelevant for modern metaphysics.''  \cite[pp. 64-65]{Cao03}}
\end{quotation}

\noindent As Einstein \cite[p. 175]{Dieks88a} made the point: ``[...] it is the purpose of theoretical physics to achieve understanding of physical reality which exists independently of the observer, and for which the distinction between `direct observable' and `not directly observable' has no ontological significance''. Observability is secondary even though ``the only decisive factor for the question whether or not to accept a particular physical theory is its empirical success.'' For the representational realist, empirical adequacy is only part of a verification procedure, not that which ``needs to be saved'' ---as van Fraassen might argue \cite{VF80}. Observability is something developed within each physical theory, it is a theoretical result, not ``a given''. As obvious as it might sound, one cannot observe a field without the notion of field, one cannot observe a dog (in general, an entity) without the notion of dog (or entity). As Einstein \cite[p. 63]{Heis71} explained to Heisenberg many years ago: ``It is only the theory which decides what we can observe.''\footnote{A very good example of how observability is developed within physical theories is Einstein's analysis of the notion of {\it simultaneity} in the context of the theory of special relativity.} 

In physical theories, it is only through metaphysical conceptual schemes ---produced through the tight interrelation of many different physical notions--- that we are capable of producing a qualitative representation and understanding of physical reality and experience. In fact, {\it Gedankenexperiments} which have many times escaped the observability of their time and adventured themselves into metaphysical debates about possible but still unperformed experiences can be only considered and imagined through the creation of adequate conceptual schemes. Indeed, as remarked by Heisenberg \cite[p. 264]{Heis73}: ``The history of physics is not only a sequence of experimental discoveries and observations, followed by their mathematical description; it is also a history of concepts. For an understanding of the phenomena the first condition is the introduction of adequate concepts. Only with the help of correct concepts can we really know what has been observed.'' 

To summarize, there are three main points which comprise the most important aspects of representational realism:

\begin{enumerate}
\item[I.] \emph{Physical Theory}: A physical theory is a mathematical formalism related to a set of physical concepts which only together provide a qualitative and quantitative understanding of a specific field of phenomena. 
\item[II.] \emph{Formal-Conceptual Representation of Reality}: Physics attempts to provide theoretical ---both formal and conceptual--- representations of physical reality. 
\item[III.] \emph{Observability is Created by the Theory}: The conditions of what is meant by `observability' are dependent and constrained by each specific theory. Observation is only possible through the development of adequate physical concepts. 
\end{enumerate}

So while empiricism considers that the world is accessible through observations, which are the key to develop scientific knowledge, our realist perspective takes the opposite standpoint and argues that it is only through the creation of theories that we are truly capable of providing understanding of our experience in the Cosmos. According to the latter view, the physical explanation of our experience goes very much against ``common sense'' observability. In fact, we could say that the history of physics is also the history of how ``common sense'' changes: it was not evident for the ancient communities that the earth is a sphere rather than a plane, it was not obvious for the contemporaries of Newton that the force that commands the movement of the moon and the planets is the same one determining the fall of an apple; it was not inescapable in the 18th Century that the strange phenomena of magnetism and electricity could be unified through the strange notion of electromagnetic field; and it was far from evident ---before Einstein--- that space and time are entangled, that objects shrink and time dilate with speed. From our viewpoint, we will only understand QM when we are capable of creating a new ``quantum sense'' which relates the mathematical equations of the theory of quanta to adequate physical notions.

\section{Aristotelian Hylomorphic Metaphysics}

The debate in Pre-Socratic philosophy is traditionally understood ---through the interpretation of both Plato and Aristotle--- as the contraposition of the Heraclitean and the Eleatic schools of thought \cite{Sambursky88}. Heraclitus was considered as defending a theory of flux, a doctrine of permanent motion and unstability in the world. He stated that the ever ongoing change or motion characterizes this world and its phenomena. This doctrine precluded, as both Plato and Aristotle, the possibility to develop certain knowledge about the world. As remarked by Verelst and Coecke \cite[p.2]{VerelstCoecke}: ``This is so because Being, over a lapse of time, has no stability. Everything that it is at this moment changes at the same time, therefore it is not. This coming together of Being and non-Being at one instant is known as the principle of coincidence of opposites.'' In contraposition to the Heraclitean school we find Parmenides as the  main character of the Eleatic school. Parmenides, as also interpreted by Plato and Aristotle, taught the non-existence of motion and change in reality. In his famous poem Parmenides stated maybe the earliest intuitive exposition of the {\it principle of non-contradiction}; i.e. that which {\it is} can only {\it be}, that which {\it is not, cannot be}. In his own turn, Aristotle developed a metaphysical scheme in which, through the notions of {\it actuality} and {\it potentiality}, he was able to articulate both the Heraclitean and the Parmenidean theories. The well known phrase of Aristotle: ``Being is said in different ways'' refers to the modes of being in which Being itself can be thought to exist in the realm of actuality and in the realm of potentiality.

The metaphysical representation or transcendent description of the world provided by Plato in the {\it Sophist} and by Aristotle in {\it Metaphysics} through specifically designed categories is considered by many to be the origin itself of metaphysical thought. In the case of Aristotle, actuality provides a metaphysical representation of a mode of existence which ---contrary to the empiricist use of the same term--- is independent of observation and the {\it hic et nunc}. This is the way through which metaphysical thought was able to go beyond the appearances of particular observations. A represented conceptual world beyond the observed world, {\it hic et nunc}. 

In relation to the stable non-contradictory realm of {\it hic et nunc} actual existence, Aristotle developed ---in order to solve the problem of movement--- a logical scheme in which the principles of {\it existence}, {\it non-contradiction} and {\it identity} would constitute within the realm of actuality the concept of actual entity.\footnote{There are three main principles which determine classical (Aristotelian) logic, namely, the existence of objects of knowledge, the principle of non-contradiction and the principle of identity. As noticed by Verelst and Coecke, these principles are ``exemplified in the three possible usages of the verb `to be': existential, predicative, and identical. The Aristotelian syllogism always starts with the affirmation of existence: something is. The principle of contradiction then concerns the way one can speak (predicate) validly about this existing object, i.e. about the true and falsehood of its having properties, not about its being in existence. The principle of identity states that the entity is identical to itself at any moment (a=a), thus granting the stability necessary to name (identify) it.'' \cite[p 167]{VerelstCoecke}.} Through these principles the notion of entity was capable of unifying, of totalizing in terms of a ``sameness''  creating certain stability for knowledge to be possible. Against Funes, it was this metaphysical architectonic which allowed us to use the same name for the dog at three-fourteen (seen in profile) and the dog at three fifteen (seen from the front). The notion of `dog' provides in this case the moment of unity of the multiple different observational phenomena. Already from this example it becomes evident that even the most common experience with a table or a chair is always metaphysically grounded. 

Apart from considering the realm of actuality as part of his metaphysical scheme, Aristotle also characterized a different realm of existence which he called `potentiality'. Furthermore, in the book $\Theta$ of {\it Metaphysics}, Aristotle [1046b5-1046b24] remarks there are two types of potentiality: ``[...] some potentialities will be non-rational and some will be accompanied by reason.'' For obvious reasons Aristotle called these two potentialities `rational' and `irrational'. In the following we shall expose in some detail these two very different notions.

\subsection{Irrational Potentiality}

In his book, {\it Potentialities}, Giorgio Agamben discusses the meaning of irrational potentiality in Aristotle's metaphysics. According to him: ``There is a generic potentiality, and this is the one that is meant when we say, for example, that a child has the potential to know, or that he or she can potentially become the head of the State.'' The child has the potentiality to become something else than what he is in actuality. Irrational potentiality implies a realm of `indefiniteness', a realm of `incompleteness' and `lack'. It is then, only when turning into actuality, that the potential is fulfilled, completed. The child becomes then a man, the seed can transform into a tree.

\begin{quotation}
\noindent{\small ``The word `actuality', which we connect with fulfillment,
has, strictly speaking, been extended from movements to other
things; for actuality in the strict sense is identified with
movement. And so people do not assign movement to non-existent
things, though they do assign some other predicates. E.g. they say
that non-existent things are objects of thought and desire, but not
that they are moved; and this because, while they do not actually
exist, they would have to exist actually if they were moved. For of
non-existent things some exist potentially; but they do not exist,
because they do not exist in fulfillment.'' Aristotle
[1047b3-1047b14]}
\end{quotation}

\noindent The path from irrational potentiality into actualization is related to the process through which {\it matter} is {\it formed}. The matter of a substance being the stuff it is composed of; the form, the way that stuff is put together so that the whole it constitutes can perform its characteristic functions. Through this passage substance becomes more perfect and, in this way, closer to God, {\it pure actus} [1051a4-1051a17].\footnote{As noticed by Verelst and Coecke \cite[p. 168]{VerelstCoecke}: ``change and motion are intrinsically not provided for in this [Aristotelian logical] framework; therefore the ontology underlying the logical system of knowledge is essentially static, and requires the introduction of a First Mover with a proper ontological status
beyond the phenomena for whose change and motion he must account
for.'' This first mover is God, {\it pure actus}, pure definiteness
and form without the contradiction and evil present in the potential
matter.} But due to this dependence, it makes no sense to consider the realm of
irrational potentiality independently of the actual realm. The final cause plays
an essential role connecting the potential and the actual realms. As noticed by Smets in
\cite{Smets05}, the idea of irrational potentiality is directly
linked to Aristotle's theory of teleological causality: ``the
transition from being [irrational] potential to actual has to be
placed within the context of [Aristotle] theory of movement and
change, which is embedded in his teleological conception of
causality [1050a7].'' It is this teleological aspect which shows the
extreme delimitation of the realm of irrational potentiality with
respect to actuality. Irrational potentiality can be only thought in
terms of its actualization, in terms of its passage into the actual.

Although Aristotle first argues that both actuality and potentiality
must be considered as independent ontological modes of existence it
becomes clear that very soon he chose the actual realm as superior
to the potential one (see \cite{Cohen09}, section 12). However,
and independently of this choice, according to Agamben
\cite[p. 179]{Agamben99}, it is not this potentiality which seems
to interests Aristotle, rather, it is ``the one that belongs to
someone who, for example, has knowledge or ability. In this sense,
we say of the architect that he or she has the {\it potential} to
build, of the poet that he or she has the {\it potential} to write
poems. It is clear that this {\it existing} potentiality differs
from the {\it generic} potentiality of the child.'' We shall now
turn our attention to this second kind of potentiality which, we
believe, can allow us to develop a notion truly independent of the actual realm and actualization ---evading at the same time teleological considerations.

\subsection{Rational Potentiality}

{\it Rational potentiality} is characterized by Aristotle as related
to the problem of possessing a capability, a faculty
[1046b5-1046b24], to what I mean when I say: ``I can'', ``I
cannot''. As explicitly noticed by Aristotle, potentiality implies a
mode of existence which must be considered as real as actuality. In
chapter 3 of book  $\Theta$ of {\it Metaphysics} Aristotle
introduces the notion of rational potentiality. In doing so Aristotle goes
against the Megarians who considered actuality as the only mode of
existence:

\begin{quotation}
\noindent{\small ``There are some who say, as the Megaric school does, that a
thing can act only when it is acting, and when it is not acting it
cannot act, e.g. he who is not building cannot build, but only he
who is building, when he is building; and so in all other cases. It
is not hard to see the absurdities that attend this view. For it is
clear that on this view a man will not be a builder unless he is
building (for to be a builder is to be able to build), and so with
the other arts. If, then, it is impossible to have such arts if one
has not at some time learnt and acquired them, and it is then
impossible not to have them if one has not sometime lost them
(either by forgetfulness or by some accident or by time; for it
cannot be by the destruction of the object itself, for that lasts
for ever), a man will not have the art when he has ceased to use it,
and yet he may immediately build again; how then will he have got
the art? [...] evidently potentiality and actuality are different;
but these views make potentiality and actuality the same, so that it
is no small thing they are seeking to annihilate. [...] Therefore it
is possible that a thing may be capable of being and not be, and
capable of not being and yet be, and similarly with the other kinds
of predicate; it may be capable of walking and yet not walk, or
capable of not walking and yet walk.'' [1046b29 -
1047a10]}
\end{quotation}

\noindent While non-rational potentialities which ``are all productive of one effect each'' rational potentialities ``produce contrary effects'' [1048a1-1048a24].  This also means that potentiality is capable of `being' and `not being'
at one and the same time: ``Every potentiality is at one and the same time a
potentiality for the opposite; for, while that which is not capable
of being present in a subject cannot be present, everything that is
capable of being may possibly not be actual. That, then, which is
capable of being may either be or not be; the same thing, then, is
capable both of being and of not being.'' [1050b7-1050b28] It is important to notice for our purposes that rational potentiality can become actual only when the state of affairs allows it. Thus, ``[...] everything which has a rational potentiality, when it desires that for which it has a potentiality and in the circumstances in which it has it, must do this. And it has the potentiality in question when the passive object is present and is in a certain state; if not it will not be able to act.'' [1048a1-1048a24] This opens the question of the contextual existence of such potentiality which might be regarded as independent (or not) of the actual state of affairs. While irrational potentialities are automatically triggered when active and passive potentialities come together, this is not the case with rational potentialities, as a rational agent can choose to withhold the realization of the potentiality even though he could realize it. This, we believe, allows us to think of a realm of potentiality completely independent of actuality. But contrary to this possible development, it seems Aristotle choose once again to limit the expressivity of potentiality within the gates of the actual realm.

\begin{quotation}
\noindent{\small ``To add the qualification `if nothing external prevents it'
is not further necessary; for it has the potentiality in so far as
this is a potentiality of acting, and it is this not in all
circumstances but on certain conditions, among which will be the
exclusion of external hindrances; for these are barred by some of
the positive qualifications. And so even if one has a rational wish,
or an appetite, to do two things or contrary things at the same
time, one cannot do them; for it is not on these terms that one has
the potentiality for them, nor is it a potentiality for doing both
at the same time, since one will do just the things which it is a
potentiality for doing.'' [1048a25-1048b9]}
\end{quotation}

From chapter 6 of book $\Theta$, Aristotle concentrates on the relation between potentiality and actuality placing actuality as the cornerstone of his architectonic, relegating potentiality to a mere supplementary role: ``We have distinguished the various senses of `prior', and it is clear that actuality is prior to potentiality. [...] For the action is the end, and the actuality is the action. Therefore even the word `actuality' is derived from
`action', and points to the fulfillment.'' [1050a17-1050a23]
Aristotle then continues to provide arguments in this line which
show ``that the good actuality is better and more valuable than the
good potentiality is evident'' [1051a4-1051a17] (see
\cite[Sect. 12]{Cohen09}).

Unfortunately, the restrictions imposed even to the notion of rational potentiality ---as related to the actual realm--- constrained the possibilities of a truly independent development of the potential realm beyond actuality. As recognized by Wolfgang Pauli:

\begin{quotation}
\noindent{\small ``Aristotle [...] created the important concept of {\small
{\it potential being}} and applied it to {\small {\it hyle}}. [...]
This is where an important differentiation in scientific thinking
came in. Aristotle's further statements on matter cannot really be
applied in physics, and it seems to me that much of the confusion in
Aristotle stems from the fact that being by far the less able
thinker, he was completely overwhelmed by Plato. He was not able to
fully carry out his intention to grasp the {\small
\emph{potential}}, and his endeavors became bogged down in early
stages.''  \cite[p. 93]{PauliJung}}
\end{quotation}

\section{The Newtonian Atomist Metaphysics of Classical Physics}

As remarked by Giorgio Agamben \cite{Agamben99}: ``The concept of potentiality has a long history in Western philosophy, in which it has occupied a central position at least since Aristotle. In both his metaphysics and physics, Aristotle opposed potentiality to actuality, {\it dynamis} to {\it energeia}, and bequeathed this opposition to Western philosophy and science.'' However, the importance of potentiality, which was first placed by Aristotle on equal footing to actuality as a mode of existence, was soon diminished in the history of Western thought. As we have seen above, it could be argued that the seed of this move was already present in the Aristotelian architectonic itself, whose focus was clearly placed in the actual realm. The realm of potentiality, as a different (ontological) mode of the being was neglected, becoming not more than a mere (logical) {\it possibility}, a process of fulfillment. In relation to development of physics, the focus and preeminence was also given to actuality. The XVII century division between {\it res cogitans} and {\it res extensa} played in this respect an important role separating also the realms of actuality and potentiality. The philosophy which was developed after Descartes kept `res cogitans' (thought) and `res extensa' (entities occupying space-time) as separated realms.\footnote{While `res cogitans', the soul, was related to the {\it indefinite} realm of potentiality and is discussed by Aristotle in {\it De Anima}, `res extensa', the entities as characterized by the principles of logic gave place to the actual considered in terms of {\it definiteness}.} As Heisenberg makes the point:

\begin{quotation}
\noindent{\small ``Descartes knew the undisputable necessity of the
connection, but philosophy and natural science in the following period developed on the basis of the polarity between the `res cogitans' and the `res extensa', and natural science concentrated its interest on the `res extensa'. The influence of the Cartesian division on human thought in the following centuries can hardly be overestimated, but it is just this division which we have to criticize later from the development of physics in our time.'' \cite[p. 73]{Heis58}}
\end{quotation}

\noindent This materialistic conception of science based itself on the main idea that extended things exist as being absolutely definite, that is, as existents within the actual realm. Paradoxically as it might seem, the division produced in the XVII century between {\it res cogitans} and {\it res extensa} together with the subsequent preeminence of ``extended things'' could be understood as the triumph of the actualist Megarian path over Aristotelian Hylomorphic metaphysics. In this respect, it is also true that the transformation from medieval to modern science coincides with the abolition of Aristotelian metaphysics as the foundation of knowledge. However, the basic structure of his metaphysical scheme and his logic still remained the basis for correct reasoning, the principle of non-contradiction ---as Kant, Leibniz and many others proclaimed--- the most certain of all principles.\footnote{As noticed by Verlest and Coecke \cite[p. 7]{VerelstCoecke}: ``Dropping Aristotelian metaphysics, while at the same time continuing to use Aristotelian logic as an empty `reasoning apparatus' implies therefore loosing the possibility to account for change and motion in whatever description of the world that is based on it. The fact that Aristotelian logic transformed during the twentieth century into different formal, axiomatic logical systems used in today's philosophy and science doesn't really matter, because the fundamental principle, and therefore the fundamental ontology, remained the same ([40], p. xix). This `emptied' logic actually contains an Eleatic ontology, that allows only for static descriptions of the world.''} 

It was Isaac Newton who was able to translate into a closed mathematical formalism both the ontological presuppositions present in Aristotelian logic and the materialistic ideal of {\it res extensa} together with actuality as its mode of existence. He did so with the aid of atomistic metaphysics. In the VI Century B.C., Leucipo and Democritus had imagined existence as consisting of small simple bodies with mass. According to their metaphysical theory, atoms were conceived as small individual substances, indivisible and separated by void. Atoms ---which means ``not divisible''--- were, for both Leucipo and Democritus, the building blocks of our material world. Many centuries later, Newton had been able not only to mathematize atoms as points in phase space, he had also constructed an equation of motion for the trajectory ---within absolute space-time-- of such ``elementary particles''. The obvious and most frightening conclusion implied by the conjunction of atomism and Newtonian's use of the {\it effective cause} was derived by Pierre Simon Laplace: 

\begin{quotation}
\noindent{\small ``We may regard the present state of the universe as the effect of its past and the cause of its future. An intellect which at a certain moment would know all forces that set nature in motion, and all positions of all items of which nature is composed, if this intellect were also vast enough to submit these data to analysis, it would embrace in a single formula the movements of the greatest bodies of the universe and those of the tiniest atom; for such an intellect nothing would be uncertain and the future just like the past would be present before its eyes.'' \cite[p. 4]{Laplace}}
\end{quotation}

\noindent The abolition of free will in the materialistic realm was the highest peak of the division between res cogitans and res extensa. In the XVII Century, in the newly proposed mechanical description of the world, the very possibility of indetermination present before in the potential realm had been erased from (physical) existence. 

In classical mechanics, every physical system may be described exclusively by means of its actual properties. A point in phase space is related to the set of values of properties that characterize the system. In fact, an actual property can be made to correspond to the set of states (points in phase space) for which this property is actual. Thus, the change of the system may be described by the change of its actual properties. Potential or possible properties are then considered as the points to which the system might (or might not) arrive in a future instant of time. Such properties are thought in terms of irrational potentiality; as properties which might possibly become actual in the future. As also noted by Dieks \cite[p. 124]{Dieks10}: ``In classical physics the most fundamental description of a physical system (a point in phase space) reflects only the actual, and nothing that is merely possible. It is true that sometimes states involving probabilities occur in classical physics: think of the probability distributions $\rho$ in statistical mechanics. But the occurrence of possibilities in such cases merely reflects our ignorance about what is actual. The statistical states do not correspond to features of the actual system (unlike the case of the quantum mechanical superpositions), but quantify our lack of knowledge of those actual features.'' Classical mechanics tells us via the equation of motion how the state of the system moves along the curve determined by initial conditions in the phase space and thus, any mechanical property may be expressed in terms of phase space variables. Needless to say, in the classical realm the measurement process plays no distinctive role and actual properties fit the definition of  {\it elements of physical reality} in the sense of the EPR paper \cite{EPR}. Moreover, the structure in which actual properties may be organized is the (Boolean) algebra of classical logic.

\section{Heisenberg's Return to Aristotelian Hylomorphism in QM}

The  mechanical description of the world provided by Newton can be
sketched in terms of static pictures which provide at each instant
of time the set of definite actual properties which constitute an actual state
of affairs (see \cite{KarakostasHadzidaki05}, p. 609). Even though the potential has been erased completely, there is in this description an obvious debt to part of the Aristotelian metaphysical scheme. The description of motion is then given, not {\it via} the path from the irrational potential to the actual, not from {\it matter} into {\it formed matter}, but rather {\it via} the successions of completely defined and determined actual states of affairs (i.e., ``pictures''
constituted by sets of actual properties with definite values). As we
discussed above, potentiality becomes then completely superfluous. 

With the advent of modern science and the introduction of mathematical
formalisms, physics seemed capable of reproducing the evolution of the
universe in a mechanical manner; just like the complicated composition of a clock allows to account for the passage of time. As Heisenberg explains, this materialistic conception of science chose actuality as the main notion to conceive existence and reality:

\begin{quotation}
\noindent{\small ``In the philosophy of Aristotle, matter was thought of in
the relation between form and matter. All that we perceive in the
world of phenomena around us is formed matter. Matter is in itself
not a reality but only a possibility, a `potentia'; it exists only
by means of form. In the natural process the `essence,' as Aristotle
calls it, passes over from mere possibility through form into
actuality. [...] Then, much later, starting from the philosophy of
Descartes, matter was primarily thought of as opposed to mind. There
were the two complementary aspects of the world, `matter' and
`mind,' or, as Descartes put it, the `res extensa' and the `res
cogitans.' Since the new methodical principles of natural science,
especially of mechanics, excluded all tracing of corporeal phenomena
back to spiritual forces, matter could be considered as a reality of
its own independent of the mind and of any supernatural powers. The
`matter' of this period is `formed matter,' the process of formation
being interpreted as a causal chain of mechanical interactions; it
has lost its connection with the vegetative soul of Aristotelian
philosophy, and therefore the dualism between matter and form
[potential and actual] is no longer relevant. It is this concept of
matter which constitutes by far the strongest component in our
present use of the word `matter'.'' \cite[p. 129]{Heis58}}
\end{quotation}

\noindent As mentioned above, in classical mechanics the mathematical
description of the behavior of a system may be formulated in terms
of the set of actual properties. The same treatment can be applied
to QM. However, the different structure of the physical properties of a quantum system imposes a deep change of nature regarding the meaning of possibility and potentiality. 

QM was related to modality since Born's interpretation of the quantum wave function $\Psi$ as a density of probability. But it was clear from the very beginning that this new quantum probability was something completely different from that considered in classical theories. ``[The] concept of the probability wave [in quantum mechanics] was something entirely new in theoretical physics since Newton. Probability in mathematics or in statistical mechanics means
a statement about our degree of knowledge of the actual situation.
In throwing dice we do not know the fine details of the motion of
our hands which determine the fall of the dice and therefore we say
that the probability for throwing a special number is just one in
six. The probability wave function, however, meant more than that;
it meant a tendency for something.'' \cite[p. 42]{Heis58} It was
Heisenberg himself who tried to interpret for the first time the wave function in terms of the Aristotelian notion of potentia. Heisenberg [{\it Op. cit.}, p. 156] argued that the concept of probability wave ``was a quantitative version of the old concept of `potentia' in Aristotelian philosophy. It introduced something standing in the middle between the idea of an event and the actual event, a strange kind of physical reality
just in the middle between possibility and reality.'' According to
him, the concept of potentiality as a mode of existence had
been used implicitly or explicitly in the development of quantum
mechanics: ``I believe that the language actually used by physicists
when they speak about atomic events produces in their minds similar
notions as the concept of `potentia'. So physicists have gradually
become accustomed to considering the electronic orbits, etc., not as
reality but rather as a kind of `potentia'.''  

But even though Heisenberg criticized the abolition of the potential realm in science and attempted to reintroduce it in order to overcome the interpretational problems of QM, when doing so he restricted potentiality ``right from the start'' to the sole consideration of {\it irrational potentiality}. As we have seen above, irrational potentiality is only subsidiary, through the teleological relation of actualization, to the actual realm; it cannot be thought to exist beyond its future actuality. Thus, by restricting his analysis of the potential realm, Heisenberg was trapped ``right from the start'' within an actualist account of reality.\footnote{In this sense it is interesting to take into account the question posed by Heisenberg to Henry Stapp regarding the ontological meaning of ideas: ``When you speak about the ideas (especially in [Section 3.4]), you always speak about human ideas, and the question arises, do these ideas `exist' outside of the human mind or only in the human mind?  In other words: Have these ideas existed at a time when no human mind existed in the world. (Heisenberg, 1972)'' \cite{Stapp09}.} We might remark that in this respect atomistic metaphysics has also played a major role constraining the possibilities of analysis. Even though the idea of QM as a theory that described atoms was severly criticized by Heisenberg and many others since the time of its construction, the language used by quantum physicists to refer to the formalism of the theory remained ---up to the present day--- an inadequate ``language of elementary particles''.

\section{Causal Powers as Future Possible Existents}

Closely related to the development of Heisenberg in terms of (irrational)
potentialities stands the development of Henri Margenau and Karl Popper in
terms of latencies, propensities or dispositions. As recalled by Mauricio
Su\'arez \cite{Suarez07}, Margenau was the first to introduce in
1954 a dispositional idea in terms of what he called {\it
latencies}. In Margenau's interpretation the probabilities are given
an objective reading and understood as describing tendencies of
latent observables to take on different values in different contexts
\cite{Margenau54}. Later, Popper \cite{Popper82}, followed by
Nicholas Maxwell \cite{Maxwell88}, proposed a propensity
interpretation of probability. Quantum reality was then
characterized by irreducibly probabilistic real propensity
(propensity waves or propensitons).\footnote{The realist position of
Popper attempted to evade the subjective aspect of Heisenberg's interpretation
\cite[p. 67-69]{Heis58} according to which: ``[The quantum] probability function combines objective and subjective elements. It contains statements on possibilities, or
better tendencies (`potentiae' in Aristotelian philosophy), and such
statements are completely objective, they don't depend on any
observer the passage from the `possible' to the real takes place
during the act of observation.''} More recently, Su\'arez has put forward a
new propensity interpretation in which the quantum propensity is
intrinsic to the quantum system and it is only the manifestation of
the property that depends on the context \cite{Suarez04a, Suarez04b,
Suarez07}. Mauro Dorato has also advanced a dispositional approach
towards the GRW theory \cite{Dorato06, Dorato10, Dorato11}. The GRW
theory after their creators: Ghirardi, Grimmini and Weber
\cite{GRW}; is a dynamical reduction model of non-relativistic
QM which modifies the linearity of Schr\"odinger's
equation. As remarked by Dorato \cite[p. 11]{Dorato06}:
``According to this reduction model, the fundamentally stochastic
nature of the localization mechanism is not grounded in any
categorical property of the quantum system: the theory at present
stage is purely `phenomenological', in the sense that no `deeper
mechanism' is provided to account for the causes of the
localization. `Spontaneous', as referred to the localization
process, therefore simply means `uncaused'.'' In \cite{Dorato06}, Dorato discusses the meaning of dispositions and reviews the need of different interpretations of QM to account for such intrinsic tendencies within the theory:

\begin{quotation}
\noindent{\small ``[...] whether and in what sense QM, in its various
interpretations, forces us to accept the existence of ungrounded,
irreducible, probabilistic dispositions, i.e. dispositions, that,
unlike fragility or permeability, lack any categorical basis to
which they can be reduced to. My claim is that the presence of
irreducible quantum dispositions in many (but not all)
interpretations involves the difficulty of giving a spatiotemporal
description to quantum phenomena, and is therefore linked to our
lack of understanding of the theory, i.e., of our lack of a clear
ontology underpinning the formalism.'' [{\it Op. cit.}, p. 3]}\end{quotation}

\noindent Dorato also explains very clearly the meaning of dispositional properties as well as their relation to categorical properties:

\begin{quotation}
\noindent{\small ``Intuitively, a disposition like permeability is not
directly observable all the times, as is the property given by the
form of an object (`being spherical'), but becomes observable only
when the entity possessing it interacts with water or other fluids.
[...] From these ordinary language examples, it would seem that the
function of dispositional terms in natural languages is to encode
useful information about the way objects around us would behave were
they subject to causal interactions with other entities (often
ourselves). This remark shows that the function of dispositional
predicates in ordinary language is essentially predictive. [...] In
a word, dispositions express, directly or indirectly, those
regularities of the world around us that enable us to predict the
future. Such a predictive function of dispositions should be
attentively kept in mind when we will discuss the `dispositional
nature' of microsystems before measurement, in particular when their
states is not an eigenstate of the relevant observable. In a word,
the use of the language of `dispositions' does not by itself point
to a clear ontology underlying the observable phenomena, but,
especially when the disposition is irreducible, refers to the
predictive regularity that phenomena manifest. Consequently,
attributing physical systems irreducible dispositions, even if one
were realist about them, may just result in more or less covert
instrumentalism.'' [{\it Op. cit.}, pp. 2-4]}\end{quotation}

\noindent In favor of dispositions, he argues [{\it Op. cit.}, p. 5] that contextuality seems to call for dispositional properties: ``Within QM, it seems natural to replace `dispositional properties' with `intrinsically indefinite properties', i.e. with properties that before measurement are objectively and actually `indefinite' (that is, without a precise, possessed value). So the passage from dispositional to non-dispositional is the passage from the indefiniteness to the definiteness of the relevant properties, due to measurements interactions.'' We can see here the direct relation between Aristotle's metaphysics, his potentiality-actuality scheme conceived in terms of causality, and the dispositional account developed in order to understand QM in terms of {\it causal capacities} \cite{Esfeld11}. It is also interesting to notice that the joint proposal of Dorato and Esfeld regarding the interpretation of QM relies on the a-causal stochastic GRW theory. Going back to dispositions and the remark of Dorato, it is interesting to notice that his idea of `observability' determines very explicitly the distinction between dispositional and categorical properties. This idea goes against our representational realist stance. But independently of our critical considerations regarding ``common sense'' observability, it is not at all clear if such dispositions are not simply a ``black box'' where we can hide the mystery surrounding QM. ``It must be granted that introducing irreducible physical dispositions is implicitly admitting that there is something we don't understand. Admitting an in-principle lack of any categorical basis to which dispositions could be reduced, in both the non-collapse views and Bohr's seems a way to surrender to mystery.'' [{\it Op. cit.}, p. 9]. As very clearly exposed by Dorato:

\begin{quotation}
\noindent{\small ``That the distinction between dispositions and categorical
properties cannot be so sharp is further confirmed by Mumford's
analysis of the problem of the reducibility of dispositions to their
so-called `categorical basis'. According to Mumford (1998), the
difference between a dispositional property like fragility and the
microscopic property of glass constituting its categorical basis is
merely linguistic, and not ontological. Referring to a property by
using a dispositional term, or by choosing its categorical-basis
terms, depends on whether we want to focus on, respectively, the
functional role of the property (the causal network with which it is
connected), or the particular way in which that role is implemented
or realized.

But notice that if we agree with Mumford's analysis, it follows that
it makes little sense to introduce irreducible quantum dispositions
as ontological hypotheses. If, by hypothesis, no categorical basis
were available, we should admit that we don't not know what we are
talking about when we talk the dispositional language in QM, quite
unlike the cases in which we refer to `fragility' or `transparency',
in which the categorical bases are available and well-known.
Introducing irreducible quantum dispositions would simply be a
black-box way of referring to the functional role of the
corresponding property, i.e., to its predictive function in the
causal network of events.

In a word, the use of the language of `dispositions' by itself does
not point to a clear ontology underlying the observable phenomena.
On the contrary, when the dispositions in question are irreducible
and their categorical bases are unknown, such a use should be
regarded as a shorthand to refer to the regularity that phenomena
manifest and that allow for a probabilistic prediction.
Consequently, {\it attributing physical systems irreducible dispositions
may just result in a more or less covert instrumentalism, unless the
process that transforms a dispositional property into a
categorically possessed one is explained in sufficient detail.}'' [{\it Op. cit.}, pp. 8-9] (emphasis added)}\end{quotation}

\noindent Dispositional proposals need thus to provide descriptions of the selecting physical process which takes place during the path from the indefinite level of dispositional properties to the definite level of actual properties. Without such explanation, the measurement problem remains unsolved. 

Causal powers have been developed in the context of QM in order to provide an understanding of the multiple terms within a superposition and provide in this way an answer to the infamous measurement problem (see \cite{deRonde17} for a detailed analysis of the measurement process in QM). On the one hand, the problem in question assumes an empiricist perspective according to which the observation of `clicks' in detectors is unproblematic and treated as a ``self evident given''. On the other hand, even though dispositionalist and propensity type interpretations go as far as claiming that propensitons and dispositions are real, they remain at the same time captive of atomistic metaphysics ---they keep holding on to the claim that ``QM talks about elementary particles''. The propensity and dispositional interpretations of QM rest thus on a paradoxical tension: on the one hand, causal powers still attempt to describe the metaphysical existence of unobservable elementary particles, but on the other hand, they define such existence only in relation to the process of actualization and the observability of `clicks' in detectors. The entanglement between the un-observable metaphysical existence of elementary particles and the observability of `clicks' in detectors as the foundation for understanding physical theories seems to create a moebius reasoning strip which rather than providing rational constraints of debate seems to be a fountain of paradoxes and pseudoproblems.

\section{Immanent Powers as Intensive Existents}

As we discussed above, representational realism assumes a very different standpoint with respect to the empiricist based characterization of physical theories. While the orthodox viewpoint in philosophy of physics continues to consider theories from an empiricist perspective according to which observability is the basis for the development of science, our neo-spinozist realist perspective returns to the original Greek understanding of physics as a discipline which through theories provide the foundation for the expression of {\it physis} in representational terms \cite{deRonde14}. In this context, our approach stresses the need to provide a conceptual representation of the mathematical formalism, one which need not be constrained or reduced to our ``common sense'' observability of chairs and tables. Since the conceptual representation of QM seems to escape the limits imposed by classical notions ---including that of ``atom'' or ``elementary particle''---, instead of insisting dogmatically to apply the metaphysical worldview inherited from Newton and Maxwell, we believe it might seem wise to start searching ---against Bohr's viewpoint\footnote{According to Bohr \cite[p. 7]{WZ}: ``[...] the unambiguous interpretation  of any measurement must be essentially framed in terms of classical physical theories, and we may say that in this sense the language of Newton and Maxwell will remain the language of physicists for all time.'' Closing the possibility of creating new physical concepts, Bohr [{\it Op. cit.}] argued that ``it would be a misconception to believe that the difficulties of the atomic theory may be evaded by eventually replacing the concepts of classical physics by new conceptual forms.''}--- for new non-classical concepts in order to make sense of both the quantum formalism and quantum phenomena. It is this non-empiricist viewpoint regarding the problem of QM which allows us to ``invert'' the measurement problem and replace it by what we call {\it the superposition problem} \cite{deRonde16e}. Let us first recall the problem we are dealing with.\\

\noindent {\it {\bf Measurement Problem:} Given a specific basis (or context),\footnote{It is important to remark that, according to this definition, both superpositions and the measurement problem are basis dependent, they can be only defined in relation to a particular basis. For a detailed analysis of this subtle but most important point see \cite{daCostadeRonde16}.} QM describes mathematically a quantum state in terms of a superposition of, in general, multiple states. Since the evolution described by QM allows us to predict that the quantum system will get entangled with the apparatus and thus its pointer positions will also become a superposition,\footnote{Given a quantum system represented by a superposition of more than one term, $\sum c_i | \alpha_i \rangle$, when in contact with an apparatus ready to measure, $|R_0 \rangle$, QM predicts that system and apparatus will become ``entangled'' in such a way that the final `system + apparatus' will be described by  $\sum c_i | \alpha_i \rangle  |R_i \rangle$. Thus, as a consequence of the quantum evolution, the pointers have also become ---like the original quantum system--- a superposition of pointers $\sum c_i |R_i \rangle$. This is why the {\it MP} can be stated as a problem only in the case the original quantum state is described by a superposition of more than one term.} the question is why do we observe a single outcome instead of a superposition of them?}\\

\noindent The measurement problem attempts to justify the observation of actual measurement outcomes, its focus concentrates on the actual realm of experience. This allows us to characterize the measurement problem as an empiricist problem which presupposes ``right from the start'' the controversial idea that actual observations are perfectly well defined for quantum phenomena. However, as we noticed above, from a representational realist stance things must be analyzed from a radically different perspective for ---as Einstein remarked--- it is only the theory which can tell us what can be observed. If we are willing to investigate the physical representation of quantum superpositions beyond classical concepts ---such as `elementary particle', `wave' or `field'--- we then need to take a completely different standpoint. Instead of trying to justify what we observe in classical terms in order to ``save the phenomena'', we need to ``invert'' the measurement problem and concentrate on the formal-conceptual level. We need to think differently, we need to ask a different question. According to our novel viewpoint, attention should be focused on the conceptual representation of the mathematical expression, not in the measurement outcomes, not in the actualization process and not in the attempt to justify experience in terms of elementary particles, tables and chairs. In short, we need to create a new physical language with concepts which are adequate to account for the structural relationships implied by the quantum formalism.\footnote{This is in no way different from what Einstein did for the Lorentz transformations in his theory of special relativity.} The solution to this problem must be provided defining new concepts in a systematic manner, beyond the reference to linguistic metaphors which make an inadequate use of concepts (e.g., the notion of atom).\\

\noindent {\it {\bf Superposition Problem:} Given a situation in which there is a quantum superposition of more than one term, $\sum c_i \ | \alpha_i \rangle$, and given the fact that each one of the terms relates through the Born rule to a meaningful physical statement, the problem is how do we conceptually represent this mathematical expression? Which is the physical concept that relates to each one of the terms in a quantum superposition?}\\
 
The new technological era we are witnessing today in quantum information processing requires that we, philosophers of QM, pay attention to the developments that are taking place. We believe that an important help could be provided by philosophers of physics who should be in charge of trying to develop a conceptual representation of quantum superpositions that allows us to think in a truly quantum mechanical manner. The first step of this project must be to recognize the inadequacy of the notion of elementary particle to account for what is going on in the quantum realm. In this respect, the superposition problem opens the possibility to discuss a physical representation of reality which goes beyond the classical atomist representation of physics. Instead of keep trying ---as we have done for almost a century--- to impose dogmatically our ``manifest image of the world'' to QM, this new realist problem allows us to reflect about possible truly non-classical solutions to the question of interpretation and understanding of quantum theory.

As the reader might already suspect, all these considerations place us also in a radically different standpoint with respect to the previous developments of potentiality in terms of causal powers ---which implicitly or explicitly constrain this mode of existence to the actual realm. The notion of potentiality we have developed in the course of our investigations \cite{deRonde11, deRonde13, deRonde15a, deRonde16a, deRonde17, RFD14} is called {\it ontological potentiality}. Contrary to the empiricist project which attempts to describe the formalism in terms of actualities, we have developed this new realm of existence in order to match the features and characteristics of the quantum formalism. This implies a path, not from a presupposed metaphysical system ---such as atomism--- to the quantum formalism, but rather from the orthodox mathematical formalism of QM to an adequate metaphysical scheme which is capable of representing in qualitative terms what the theory is really talking about. 

Our proposal begins with the definition of a mode of existence, ontological potentiality, completely independent of the actual realm. It continues by defining two key notions, namely, immanent power and potentia. According to representational realism being is said in many different ways,\footnote{There is in our neo-spinozist account an implicit ontological pluralism of {\it multiple representations} which can be related to {\it one reality} through a {\it univocity principle}. This is done in analogous manner to how Spinoza considers in his immanent metaphysics the {\it multiple attributes} as being expressions of the same {\it one single substance}, namely, nature (see \cite{deRonde14, deRonde16b}). Our non-reductionistic answer to the problem of inter-theory relation escapes in this way the requirement present in almost all interpretations of QM which implicitly or explicitly attempt to explain the formalism in substantialist atomistic terms (see \cite{deRondeFM17}).} and just like particles, fields and waves are existents within the actual realm and represented by our classical theories, {\it immanent powers} with definite {\it potentia} are existents within the potential realm which require a quantum mechanical description. Our physical representation of QM can be condensed in the following seven postulates which contain the relation between our proposed new physical concepts and the orthodox formalism of the theory.

\begin{enumerate}

{\bf \item[I.] Hilbert Space:} QM is mathematically represented in a vector Hilbert space.

{\bf \item[II.] Potential State of Affairs (PSA):} A specific vector $\Psi$ with no given mathematical representation (basis) in Hilbert space represents a PSA; i.e., the definite potential existence of a multiplicity of {\it immanent powers}, each one of them with a specific {\it potentia}.

{\bf \item[III.] Quantum Situations, Immanent Powers and Potentia:} Given a PSA, $\Psi$, and the context or basis, we call a quantum situation to any superposition of one or more than one power. In general given the basis $B= \{ | \alpha_i \rangle \}$ the quantum situation $QS_{\Psi, B}$ is represented by the following superposition of immanent powers:
\begin{equation}
c_{1} | \alpha_{1} \rangle + c_{2} | \alpha_{2} \rangle + ... + c_{n} | \alpha_{n} \rangle
\end{equation}

\noindent We write the quantum situation of the PSA, $\Psi$, in the context $B$ in terms of the order pair given by the elements of the basis and the coordinates in square modulus of the PSA in that basis:
\begin{equation}
QS_{\Psi, B} = (| \alpha_{i} \rangle, |c_{i}|^2)
\end{equation}

\noindent The elements of the basis, $| \alpha_{i} \rangle$, are interpreted in terms of {\it powers}. The coordinates of the elements of the basis in square modulus, $|c_{i}|^2$, are interpreted as the {\it potentia} of the power $| \alpha_{i} \rangle$, respectively. Given the PSA and the context, the quantum situation, $QS_{\Psi, B}$, is univocally determined in terms of a set of powers and their respective potentia. (Notice that in contradistinction with the notion of {\it quantum state} the definition of a {\it quantum situation} is basis dependent and thus intrinsically contextual.)

{\bf \item[IV.] Elementary Process:} In QM we only observe in the actual realm discrete shifts of energy (quantum postulate). These discrete shifts are interpreted in terms of {\it elementary processes} which produce actual effectuations. An elementary process is the path which undertakes a power from the potential realm to its actual effectuation. This path is governed by the {\it immanent cause} which allows the power to remain potentially preexistent within the potential realm independently of its actual effectuation. Each power $| \alpha_{i} \rangle$ is univocally related to an elementary process represented by the projection operator $P_{\alpha_{i}} = | \alpha_{i} \rangle \langle \alpha_{i} |$.

{\bf \item[V.] Actual Effectuation of an Immanent Power (Measurement):} Immanent powers exist in the mode of being of ontological potentiality. An {\it actual effectuation} is the expression of a specific power within actuality. Different actual effectuations expose the different powers of a given $QS$. In order to learn about a specific PSA (constituted by a set of powers and their potentia) we must measure repeatedly the actual effectuations of each power exposed in the laboratory. (Notice that we consider a laboratory as constituted by the set of all possible experimental arrangements that can be related to the same $\Psi$.) An actual effectuation does not change in any way the PSA. 

{\bf \item[VI.] Potentia (Born Rule):} A {\it potentia} is the intensity of an immanent power to exist (in ontological terms) in the potential realm and the possibility to express itself (in epistemic terms) in the actual realm. Given a PSA, the potentia is represented via the Born rule. The potentia $p_{i}$ of the immanent power $| \alpha_{i} \rangle$ in the specific PSA, $\Psi$, is given by:
\begin{equation}
Potentia \ (| \alpha_{i} \rangle, \Psi) = \langle \Psi | P_{\alpha_{i}} | \Psi \rangle = Tr[P_{ \Psi} P_{\alpha_{i}}]
\end{equation}

\noindent In order to learn about a $QS$ we must observe not only its powers (which are exposed in actuality through actual effectuations) but we must also measure the potentia of each respective power. In order to measure the potentia of each power we need to expose the $QS$ statistically through a repeated series of observations. The potentia, given by the Born rule, coincides with the probability frequency of repeated measurements when the number of observations goes to infinity.

{\bf \item[VII.]  Potential Effectuations of Immanent Powers (Schr\"odinger Evolution):} Given a PSA, $\Psi$, powers and potentia evolve deterministically, independently of actual effectuations, producing {\it potential effectuations} according to the following unitary transformation:
\begin{equation}
i \hbar \frac{d}{dt} | \Psi (t) \rangle = H | \Psi (t) \rangle
\end{equation}
\noindent While {\it potential effectuations} evolve according to the Schr\"odinger equation, {\it actual effectuations} are particular expressions of each power (that constitutes the PSA, $\Psi$) in the actual realm. The ratio of such expressions in actuality is determined by the potentia of each power.
\end{enumerate}

\smallskip 

\noindent Let us now continue to analyze in more detail some important aspects of our interpretation:\smallskip

{\it The potential state of affairs as a sets of immanent powers with definite potentia.} Our choice to develop an ontological realm of potentiality absolutely independent of the actual realm of existence implies obviously the need to characterize this realm in an independent manner to classical physical concepts such as `particles', `waves' and `fields' ---notions which are defined in strict relation to the actual mode of existence. According to our viewpoint, while classical physics talks about systems with definite properties (`particles', `waves' and `fields'), QM talks about the existence of powers with definite potentia. While the classical representation of sets of systems with definite properties can be subsumed under the notion of an {\it actual state of affairs}, QM provides a representation in terms of a {\it potential state of affairs}. This representation seeks on the one hand, to define concepts in a systematic categorical manner avoiding metaphorical discourse, and on the other hand, to understand QM ---what it really talks about, the experience it implies--- through these new concepts in an intuitive manner.\footnote{This is what some of the founding fathers of QM understood as the {\it anschaulich} content of the theory.}  Several examples have been already discussed in \cite{deRonde16a}. In other words, we need to create a new way of thinking, with new concepts which allow us to define clearly what is observable according to the theory of quanta.  

\smallskip

{\it The existence and interaction of quantum possibilities.} The need to consider quantum possibilities as part of physical reality is supported, in the first place, by the fact that quantum probability resists an ``ignorance interpretation''. The fact that the quantum formalism implies a non-Kolmogorovian probability model which is not interpretable in epistemic terms is a well known fact within the foundational literature since Born's interpretation of the quantum wave function \cite{Redei12}.\footnote{It is true that QBism does provide a subjectivist interpretation of probability following the Bayesian viewpoint, however, this is done so at the price of denying the very need of an interpretation for QM. See for a detailed analysis: \cite{deRonde16a, deRonde16c}.} But more importantly, the quantum mechanical formalism implies that projection operators can be understood as {\it interacting} and {\it evolving} \cite{RFD14}. In classical mechanics the mathematical and a conceptual levels are interrelated in such a consistent manner that it makes perfect sense to relate mathematical equations with physical concepts. For example, the mathematical account of a point in phase space which evolves according to Newton's equation of motion can be consistently related to the trajectory of a particle in absolute space-time. Let us remark against a common naive misunderstanding between mathematics and physics: a point is not a particle, phase space is not Newtonian space-time. But in QM, while the interaction and evolution of projection operators is represented quantitively through the mathematical formalism we still lack a conceptual qualitative representation of what projection operators really mean. In this respect, the {\it interaction} in terms of entanglement, the {\it evolution} in terms of the Schr\"odinger equation of motion and the {\it prediction} of quantum possibilities in statistical terms through the Born rule are maybe the most important features which point in the direction of developing an ontological idea of possibility which is truly independent of actuality. This development is not a mathematical one; rather, it is a metaphysical or conceptual enterprise. 

\smallskip

{\it The intensity of quantum possibilities.} Another important consequence of the ontological perspective towards quantum possibilities relates to the need of reconsidering the binary existencial characterization of properties in terms of an homomorphic relation to the binary Boolean elements $\{ 0,1 \}$ (or truth tables). In \cite{deRonde16a} we proposed to extend the notion of {\it element of physical reality} escaping the characterization of existence in terms of certitude (probability = 1) and considering ``right from the start'' the quantum probabilistic measure in objective terms. This move implies the development of existence beyond the gates of certitude and the complementary need of characterizing the basic elements of our ontology ---namely, immanent powers--- in intensive terms; i.e. as relating to a value which pertains to the interval $[0,1]$. In this way, each {\it immanent power} has an intensive characterization which we call {\it potentia}. We could say that, unlike properties that pertain to systems either exist or do not exist (i.e., they are related either to 1 or 0), immanent powers have a more complex characterization which requires, apart form its binary relation to existence, a number pertaining to the closed interval $[0,1]$ which specifies its (potential) existence in an intensive manner. It is through the introduction of an intensive mode of existence that we can understand quantum probability as describing an objective feature of QM and at the same time restore its epistemic role as a way to gain knowledge about a still unknown but yet existent (potential) state of affairs.

\smallskip 

{\it Immanent powers and contextuality.} It is important to notice that the intensive characterization of immanent powers allows us to escape Kochen-Specker contextuality \cite{deRonde17b} and restore a global valuation to all projection operators of a quantum state, $\Psi$. By removing the actualist binary reference of classical properties to $\{ 0,1 \}$, and implementing instead an intensive valuation of projection operators to $[0,1]$ we are able, not only to bypass Kochen-Specker theorem \cite{KS}, but also to restore ---through a {\it global intensive valuation}--- an objective representation of the elements the theory talks about. Powers are non-contextual existents which can be defined univocally and globally for any given quantum state $\Psi$. In this way, just like in the case of classical physics, quantum contextuality can be understood as exposing the epistemic incompatibility of measurement situations and outcomes (see for a detailed discussion and analysis: \cite{deRonde16a, deRonde16d, deRonde17b}).

\smallskip 

{\it The contradiction of quantum possibilities.} Some quantum superpositions of the ``Schr\"odinger cat type'' \cite{Schr35} constituted by two contradictory terms, e.g. `$| + \rangle$' and `$| - \rangle$', present a difficult problem for those who attempt to describe the theory in terms of particles with definite non-contradictory properties. Indeed, as discussed in \cite{daCostadeRonde13}, while the first term might relate to the statement `the atom possesses the property of being decayed' the second term might relate to the statement `the atom possesses the property of not being decayed'.  Obviously, an atom cannot be `decayed' and `not decayed' at the same time ---just like a cat cannot be `dead' and `alive' simultaneously. Any physical object ---an atom, a cat, a table or a chair---, by definition, cannot posses  contradictory properties. Physical objects have been always ---implicitly or explicitly--- defined since Aristotle's metaphysics and logic in terms of the principles of existence, non-contradiction and identity. However, regardless of the manner in which objects are defined in classical physics, QM allows us to predict through the mathematical formalism how these terms will interact and evolve in different situations. The realist attitude is of course to consider that the formalism, and in particular quantum superpositions, are telling us something very specific about physical reality. It is in fact this belief which has allowed us to enter the new technological era of quantum information processing. This is also why one might consider Schr\"odinger's analysis as an {\it ad absurdum} proof of the impossibility to describe quantum superpositions in terms of classical notions (i.e., particles, waves, tables or cats). The escape road proposed by some modal interpretations \cite{Dieks07,Dieks10}  and some readings of the many words interpretation \cite{Sudbery16, Wallace07}, which attempt to consider the terms of a superposition as {\it possible future actualizations} misses the point, since the question is not the epistemic prediction or justification of future outcomes, but the understanding of what is really going on even before the measurement process has taken place. Finally, recalling that the interpretation of probability in epistemic terms is untenable within the orthodox formalism ---which is only consistent with a non-Kolmogorovian probability measure--- there seems to be no escape ---at least for a realist which attempts to be consistent with the orthodox formalism--- but to confront the fact that classical notions such as `atom', `wave', `table' or `chair' (i.e., notions categorically constrained by the principles of existence, non-contradiction and identity) are not adequate concepts to account for quantum superpositions \cite{deRonde16e}. In this respect, using a term created by Gaston de Bachelard we might say that the notion of classical entity rather than helping us to understand the quantum formalism has always played the role of an {\it epistemic obstacle} \cite{deRondeBontems11}.

\smallskip  

{\it The relation and independence of immanent powers with respect to the actual realm.} Immanent powers have an independent potential existence with respect to the actual realm. Measurement outcomes are not what potential powers attempt to describe. It is exactly the other way around ---at least for a representational realist. For the realist, measurement outcomes are only expressions of a deeper moment of unity which requires a categorical definition. This is completely analogous to the classical case in which the view of a dog at three-fourteen (seen in profile) or the dog at three fifteen (seen from the front) are both particular expressions which find their moment of unity in the notion of `dog'. However, we still need to provide an answer to the measurement problem and explain in which manner quantum superpositions (in formal terms), and powers with definite potentia (in conceptual terms), relate to actual effectuations. In \cite{deRonde17}, we have provided a detailed analysis of our understanding of how the measurement process should be understood in QM. Within our approach, the quantum measurement process is modeled in terms of the spinozist notion of {\it immanent causality}. The immanent cause allows for the expression of effects remaining both in the effects and its cause. It does not only remain in itself in order to produce, but also, that which it produces stays within. Thus, in its production of actual effects the potential does not deteriorate by becoming actual ---as in the case of the hylomorphic scheme of causal powers (see section 1, p. 4 of \cite{deRonde17}).\footnote{For a more detailed discussion of the notion of immanent cause we refer to \cite[Chapter 2]{Melamed}.} Immanent powers produce, apart from actual effectuations, also {\it potential effectuations} which take place within potentiality and remain independent of what happens in the actual realm. Within our model of measurement, while potential effectuations describe the ontological interactions between immanent powers and their potentia ---something known today as {\it entanglement}---, actual effectuations are only epistemic expressions of the potentia of powers. Actualities are only partial expressions of powers. Just like when observing a dog, a table or a chair we only see a partial perspective of the object ---the perceptual adumbration of an object in the phenomenological sense--- but never the object itself, measurement outcomes expose only a partial account of the potentia of powers.

\smallskip

{\it Relational definition of powers and their potentia.} Against the (classical) substantialist atomist representation through which most present interpretations ---implicitly or explicitly--- attempt to understand QM, our proposal attempts to consider an ontological relational scheme which understands that QM talks, rather than about ``elementary particles'' (independent substances), about relational existents, namely, immanent intensive powers (see \cite{deRondeFM17}).

\section{Final Remarks: Causal vs Immanent Powers}

As we have seen, there are many differences between causal powers and immanent powers. While causal powers are understood ---following the empiricist conservative agenda which attempts to ``save the phenomena''--- as properties that attempt to justify the appearances of observed actualities, immanent powers attempt ---following the representational realist more ambitious program--- to provide a conceptual representation and intuitive understanding of what is going on beyond measurement outcomes. While causal powers are understood as potential properties of elementary particles which at some point acquire a definite value through their actualization and interaction with the environment, immanent powers are understood as characterized in terms of a specific potentia which allows ---through the generalization of reality in intensive terms--- to restore a {\it global intensive valuation}. In turn, it is the possibility of such global valuation which ---escaping KS contextuality--- allows us to define an objective account of physical reality. While the measurement process still remains a problem within the hylomorphic metaphysics proposed by causal powers, immanent causality implements a novel manner of understanding the process of measurement in QM. It is in this way that immanent powers are able to provide an explanation of actual effectuations without invoking the ``collapse'' of the quantum wave function or turning possibilities into actualities ---as it is the case of Everett original ``Megarian interpretation'' (see \cite{daCostadeRonde13}) and some versions of the modal interpretation \cite{Dieks07}. Furthermore, while in our approach the reference to the actual realm becomes merely epistemic, it is potential effectuations which become the basis for considering the experience QM is really talking about. In this respect, our intensive approach has clear empirical differences with respect to the causal hylomorphic approaches which make use of causal powers. Immanent powers imply the existence of potential effectuations, a type of experience that we must try to understand beyond the classical actual realm. Actual effectuations are just a way to grasp potential powers when related to actual effectuations. We believe that the differences we have discussed between the orthodox notion of causal power and our proposed immanent power might allow us to better understand and even develop the theory of quanta.   

Our theory of intensive immanent powers discusses a new realm of existence which goes beyond visual observability and reconsiders the main problem of contemporary physics, that is, the need to account for a new representation of reality. As remarked by Wolfgang Pauli: 

\begin{quotation}
\noindent {\small ``When the layman says `reality' he usually thinks that he
is speaking about something which is self-evidently known; while to
me it appears to be specifically the most important and extremely
difficult task of our time to work on the elaboration of a new idea
of reality.'' \cite[p. 193]{Laurikainen98}}
\end{quotation}

\section*{Acknowledgements} 

I want to thank Ruth Kastner and Matias Graffigna for discussions on related subjects presented in this manuscript. This work was partially supported by the following grants: FWO project G.0405.08 and FWO-research community W0.030.06. CONICET RES. 4541-12 (2013-2014) and the Project PIO-CONICET-UNAJ (15520150100008CO) ``Quantum Superpositions in Quantum Information Processing''.

\end{document}